\documentclass{sig-alternate}
\usepackage{xcolor}
\usepackage{fixme}
\usepackage{flushend}

\usepackage{algorithm}
\usepackage{algpseudocode}
\fxsetup{
    status=final,
    author={Note},
    layout=footnote, % pdfnote
    theme=color,
    silent=true,
    inline=false
}
\newcommand\todo[1]{}
\begin{document}

\newcommand\rev[1]{\textcolor{black}{#1}}

% Redefine paragraph command
% 
\renewcommand\paragraph[1]{%
\vskip 2pt\noindent
\textbf{#1.}%
}

\title{Evolutionary Data Systems}

\author{
Stratos Idreos\hspace{2em} Lukas M. Maas\hspace{2em} Mike S. Kester\vspace{2em}\\
Harvard University
}

\date{\today}

\newcommand\testmachine{8-Core 2.7 Ghz Intel Xeon E5 with 64GB of main memory}

\maketitle

\begin{abstract}
Anyone in need of a data system today is confronted with numerous complex options in terms of system architectures, 
such as traditional relational databases, NoSQL and NewSQL solutions as well as several 
sub-categories like column-stores, row-stores etc.
This overwhelming array of choices makes bootstrapping data-driven applications difficult and time consuming, requiring expertise often not accessible due to cost issues (e.g., to scientific labs or small businesses).
In this paper, we present the vision of evolutionary data systems that free 
systems architects and application designers from the complex, cumbersome and expensive 
process of designing and tuning specialized data system architectures that fit only a single,
static application scenario.
Setting up an evolutionary system is as simple as identifying the data.
As new data and queries come in, the system automatically evolves so 
that its architecture matches the properties of the incoming workload at all times.
Inspired by the theory of evolution, at any given point in time, an evolutionary system 
may employ multiple competing solutions down at the low level of database architectures -- characterized as combinations of data layouts, access methods and execution strategies.
Over time, ``the fittest wins'' and becomes the dominant architecture until the environment (workload) changes.
In our initial prototype, we demonstrate solutions that can seamlessly evolve (back and forth) between a key-value store and a column-store architecture in order to adapt to changing workloads.
\end{abstract}

\section{Introduction}
\label{intro}

\paragraph{Modern Data Systems: Numerous Options - No Single Architecture Works}
The most characteristic aspect of a data system is its architecture, 
which is defined by the way it stores, accesses and processes data. 
Over the past five decades the predominant data model has been the relational model \cite{codd},
while the predominant systems architecture has been what we now call row-stores, 
i.e., systems that logically organize data in a tabular form and physically store 
and process data one row of a table at a time \cite{rowstoresArchitecture}. 
In the past few years, though, the situation has changed drastically \cite{cstoreArchitecture}. 

Due to advances in technology, we now see a plethora of new data-driven applications 
that have different requirements compared to traditional applications. 
As a response, the research community and industry have designed a plethora of 
new data systems architectures to match the new requirements. 

\vspace{.5em}
\emph{
A fundamental problem is that there is no single data systems architecture that fits the ever-increasing kinds
of data-driven applications and scenarios.
}
\vspace{.5em}

When in need for a data system, a business, a research lab or a government organization is confronted with
an array of complex options. 
For example, one may use a NoSQL solution, a NewSQL solution or a traditional SQL solution. 
In addition, within each category there are several variations to 
how we store and access data (e.g., row-stores, column-stores and hybrids in the SQL solution space). 
Clearly ``one size does not fit all'' when it comes to data 
systems architectures \cite{OneSizeFitsAllIdeaWhoseTimeHasComeAndGone,Athanassoulis2016b}, but at the same time
designing a new data system from scratch every time we need to support a new application feature, 
is not a scalable solution in the long-term; we do not live in a static world anymore in terms of application scenarios.

\paragraph{Manual Design and Tuning in Dynamic Environments}
Today we see modern businesses struggling 
to find a system architecture that fits their needs, ending up relying on multiple custom solutions 
that are difficult and expensive to deploy, tune and maintain.
For example, Facebook uses nearly a dozen different data systems architectures which are all slight variations
of key-value stores or relational systems, yet none of those two global architecture designs 
fits their needs exactly \cite{BigDataPanelSigmod14}.  
In practice, what happens is that businesses manually strip down all the features they do not need or they 
slightly alter other features to tailor-fit them to their needs, essentially manually implementing the vision of 
RISC-style database architectures \cite{RISCarchitectures}. 
This way, the resulting system has just enough footprint to avoid all overheads 
due to system components or functionalities which are not necessary for the specific task that it is used for. 
This may be OK (for now) for giant organizations but in 
typical-size businesses there is usually a shortage of (financial) resources to invest 
in designing and maintaining tailored systems to fit the ever evolving application needs.
The same argument holds in scientific research, which becomes more and more data-driven. In such contexts, support for tailored systems is as limited as resources in general. At the same time, however,   
scientific data management needs evolve continuously
\fxnote{The same argument holds in scientific research that becomes more and more data-driven
but with limited support for tailored systems and with limited resources, while at the same time   
scientific data management needs evolve continuously}
%\cite{ResearcherGuide,ScientificDataManagementGray,ManagingScientificData,RequirementsSciDB}. 
\cite{ScientificDataManagementGray}. 
To make things worse, many new kinds of data-driven applications 
are characterized by varying workload patterns and data exploration scenarios~\cite{BDE,Idreos2015,Athanassoulis2016b},
which means that no single architecture works optimally and that we cannot always anticipate 
what a good architecture is,\fxnote{added ,} as we may not have enough workload knowledge. 

\paragraph{Evolutionary Data Systems}
%The problems and scenarios discussed above create a setting where 
In this way,
data systems architectures become a fundamental bottleneck that will only become worse
as we scale our ability to collect data and organically grow new data-driven applications. 

In this paper, we ``attack'' the problem at its very core.
We present a vision for next generation data systems that are not static and monolithic in 
how they store, access and process data. 
The insight is that queries (and in turn applications) 
should be able to ``organically define''
the architecture of the data system they need in the same way that new applications are organically developed as new technology,
consumer, scientific and business needs evolve.  

Inspired by the theory of evolution and past work on adaptive database architectures, 
we present a vision for a new class of database architectures, \emph{evolutionary data systems}. 
%which evolve continuously to match the application's needs. 
An evolutionary system is not static; it blurs the boundaries between prevalent monolithic systems architectures,
NoSQL, NewSQL, SQL as well as their sub-categories, e.g., row-stores, column-stores and hybrids. 
Instead of having to ``lock'' an application to a single system architecture, an evolutionary system 
automatically ``takes the shape'' of the data and queries of a given application and continues to evolve with them.

\vspace{.5em}
\emph{All one needs to do to use an evolutionary data system is point to the data and start querying.}
\vspace{.5em}

At any point in time, multiple competing solutions regarding 
how to store and access data co-exist at the core of an evolutionary system
and then the ``fittest wins'' 
while the  system continuously and autonomously mutates, spinning off and testing new combinations of storage and access patterns
as the workload evolves.

\paragraph{Contributions and Outline}
In the rest of the paper, we discuss the vision of evolutionary systems in detail and present 
a first design of an evolutionary system that can seamlessly shift between various 
storage models such as key-value stores, column-stores and hybrid systems (Section \ref{sec:evolution}).
Through an early prototype system, we demonstrate that it is indeed feasible to 
autonomously transition between different architectures (Section \ref{sec:evaluation}). 
We then present a detailed road-map of the research challenges towards the full vision of evolutionary systems 
(Section \ref{sec:research-challenges}) 
and we position evolutionary systems with respect to pioneering  
past work from the database community that has inspired our vision (Section \ref{sec:related-work}).

\section{Evolutionary Data Systems}
\label{sec:evolution}
This section presents the core concepts of our vision and provides initial ideas on how to tackle 
some of the most challenging problems on our way to making evolutionary systems a reality, 
including how systems evolve, what happens when the workload changes, 
and how the feature set of an evolutionary system can be extended to spin off new architectures.

\subsection{The Vision}
\paragraph{Goals and Motivation}
The goal of an evolutionary system is to significantly simplify and therefore reduce 
the cost of data system design and tuning efforts. Using a data system should be 
as simple as pointing to the data and start posing queries without worrying about the 
underlying design details of storage and access patterns.

\vspace{.4em}
\emph{The architecture of a data system should not be a contract carved in stone, locking an application in
a single system architecture that may soon become suboptimal.}
\vspace{.4em}

When the workload or the requirements of the application change, we should not need to 
go through a painful and expensive process of migrating to a new system.
Migrating to a new system implies at least the following complex steps: 
(a) loading all data from scratch into the new system and its internal data layout,
and (b) tuning the new system with the proper indexes, views and for all its performance knobs.
These steps require expertise and idle time; due to the numerous complex design choices, 
even for small applications it may take several weeks or months
to reach a good system set up.

\paragraph{Continuous Evolution}
Evolutionary systems remove all complex design and tuning steps
when setting up a data system, and replace them with a continuous process of \textit{mutations} and \textit{natural selection}.
Instead of having to lock an application to a single system, an evolutionary system automatically ``takes the shape'' of the data and queries of a given application and continues to evolve with them.

The design of low-level internals of evolutionary data systems is inspired (a) by 
past work of the database community on adaptive data processing \emph{to perform lightweight adaptation actions
at the core of the data systems architecture} 
and (b) by the theory of evolution; 
At any given point in time, an evolutionary system may employ multiple competing solutions down 
at the low level of database architectures, e.g., 
various different data layouts, access methods and execution strategies. 
In addition, the system continuously 
and seamlessly mutates those solutions as the workload changes, favoring the one performing best.
In this way, instead of locking an application with a priori, complex and uncertain decisions to a given architecture,
multiple competing architectures exist in parallel.  
Over time, ``the fittest solution wins'' and becomes the dominant architecture until the environment (workload) changes.

The process of evolution is not monolithic; a single application may end up being served by multiple 
architectures depending on the query patterns;  Different queries may have different properties and may 
benefit by different layout and access methods. 
Thus, at any given point in time, part of the underlying
data may be stored in multiple formats to serve the various architecture instances necessary for the workload.
Based on this combination of randomness (\textit{mutations}), performance evaluation (\textit{selection} of the fittest solution) and recombination of existing solutions (\textit{crossover}), potentially, every variation of the architecture, small or big,  can form a candidate solution for a given workload.
The more queries arrive, the more the system evolves to reach a state
where the given workload pattern can be served by a nearly optimal architecture. 

\label{sec:architecture}
\begin{figure*}[t]
\centering
\includegraphics[width=\textwidth]{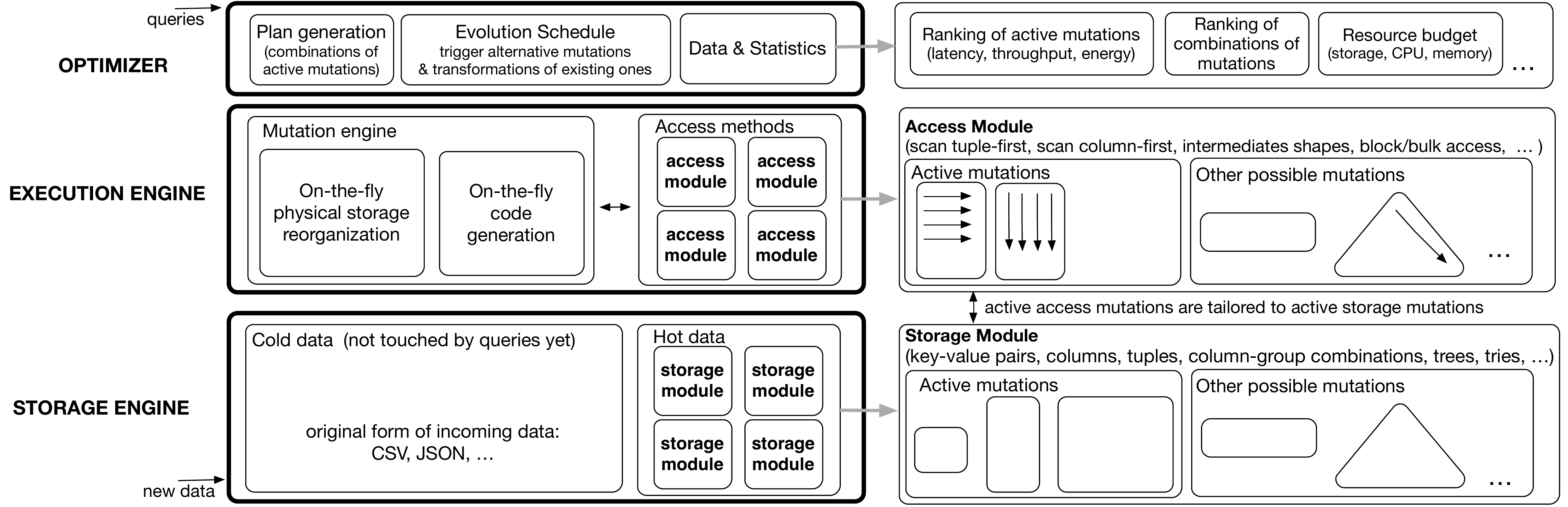}
\caption{The architecture of an evolutionary data system.}
\label{f:architecture}
\end{figure*}

\paragraph{Opportunities}
Evolutionary data systems open up an exciting new research area that touches
every corner of data systems architecture design and bring numerous opportunities to effortlessly map application
needs to tailored data systems architectures. 
The agility to test any combination of system components
allows the design of truly tailored systems at the micro-level which is hard 
to do in a manual way due to the vast number of possible combinations. 
Analysis speed and system footprint are significantly benefited. 
In this way, we can also minimize the costs associated with owning a tailored data system
which will benefit sciences, small businesses as well as big businesses with dynamic workload patterns and evolving 
requirements.

\subsection{Architecture}

The architecture of an evolutionary system should be flexible  
to be able to morph based on the characteristics of the workload.

\paragraph{Modular Architecture}
To make it possible for an evolutionary system to transition between architectures, we propose a flexible, modular architecture that is made up of sets of exchangeable components as shown in Figure~\ref{f:architecture}.
This is possible as, down at the low levels of system design, most architectures only differ slightly. 
For example, a storage module transitions from a key-value format to a pure column-store format by breaking the ``value'' part of each key-value pair down to its individual properties.
Each module comes with a set of mutation rules that indicate its operation modes.
Using these rules, the system can create and test alternative architectures by combining the various 
modules in their different mutations, 
while a system designer can extend the system by adding mutation rules to existing modules 
or whole new modules. 
Figure~\ref{f:architecture} shows examples of such modules and how they fit in the global architecture;
we discuss about modules in more detail below.

\paragraph{Storage Modules} A storage module is responsible for a specific data part 
and initiates mutations if and only if the respective data is touched by one or more queries.
Storage modules mutate with a goal to reach physical layout transformations that co-locate data which is used together in queries (and to avoid mixing it with data
which is not touched by the same set of queries). This minimizes data transfers through the memory hierarchy.
As shown in the lower right part of Figure \ref{f:architecture}, a single storage module may have multiple
active mutations (storage layouts in this case) for the same data parts depending on the requested access 
patterns, while more candidate mutations are possible and are waiting to be tested. 

\paragraph{Storage Engine}
The storage engine is responsible for managing the underlying data structures and decides, based on the workload, how to store data for different candidate solutions.
Essentially, the storage engine is a collection of storage modules, each one working independently 
to optimize storage patterns for its data. 
Unless the storage engine is explicitly initialized (e.g., by providing a set of queries), data will be kept in it's original format until it is needed.
This might be flat CSV files, JSON, XML or any other format.
The total number of modules in a storage engine may vary depending on the workload. 
In a vanilla engine, the first storage modules are born when the first queries arrive.
These storage modules contain the requested data only. 
As the workload and the system evolve,
any  storage module may (recursively) split into two new modules if the access patterns required 
for parts of its data diverge or it may duplicate itself (essentially replicating the respective data) 
to support different storage patterns for the same data at the same time. In addition, as new queries access\fxnote{arrive 
on} previously untouched data, new storage modules are created to provide support for these data parts.

\paragraph{Access Modules} An access module operates on top of a storage module 
to perform specific tasks (scans, aggregations, etc.) on the data
and can take different shapes depending on the workload. 
For example, an access module performing a scan-select may mutate in several ways: 
it may or may not use predication, it may spill out intermediate results in a tuple format, in a columnar format or in some hybrid format, it may decide to process only a small batch of data at a time, or consume its whole input in one go before pushing results to the next component in the query plan, etc. 
An example of an access module is shown in the middle right part of Figure~\ref{f:architecture}.
Similarly to storage modules, several active mutations (access patterns in this case) 
co-exist and compete in an access module, while
several more possible mutations are waiting to be tested.

\paragraph{Execution Engine}
The execution engine manages the available access modules, and enables seamless transformations between mutations of these modules via an internal mutation engine.
At any given time, multiple variations of the same access module might exist to serve different parts of the data.
To minimize the system's footprint, mutations of access modules are created on demand via on-the-fly code generation and linking.

\paragraph{Evolutionary Optimizer}
An evolutionary system still needs a component to make certain decisions and schedule mutations\fxnote{scheduling}.
An evolutionary optimizer plays this role and is responsible for prioritizing 
the choices we discussed in the previous paragraphs. 
There are a number of differences though compared to a traditional optimizer in state-of-the-art data systems.
An evolutionary\fxnote{Here an} optimizer does not need to be a single central component that has a holistic view and knowledge about the 
system. For example, each machine employed can locally perform its own optimization and spin off new architectures
if this may create better results. In addition, an evolutionary optimizer does not necessarily need to come up 
with a single choice believed to be optimal, as it can always spin off new architectures and let them 
``fight'' if the decision is a tight one. 

\begin{figure*}[t]
\centering
\includegraphics[width=.98\textwidth]{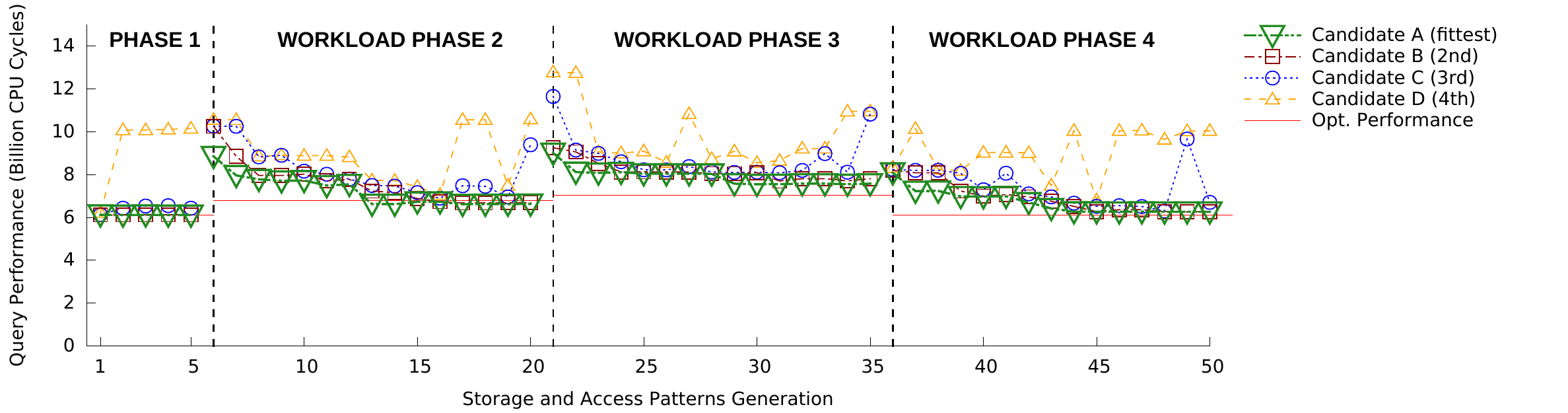}
\caption{Evolving Across Architectures.}
\label{f:exp1}
\end{figure*}

\paragraph{Survival of the Fittest}
In order to evolve the system design to better fit a given workload, an evolutionary system continuously generates and evaluates alternative architecture and design solutions. Architectures that show more promising performance results are favored and function as blueprints for new architectures that replace less performant solutions.
The metric that defines which solution is best can be a configurable parameter as different applications
have different requirements. The standard metric is response time for individual queries, latency,
and throughput for a sequence of queries (how many queries can the system serve within a time unit).
However, the evolution process is orthogonal to the specific metric.
For example, certain applications might be willing to sacrifice
a bit in terms of response time or throughput, if that translates to significant savings in energy costs or they might be willing to operate only within specific energy cost boundaries. 
An interesting approach is to incorporate application-level feedback to rate the behavior of specific requests/queries;
the evolutionary data system can then use this feedback to rank the candidate solutions.

\paragraph{Managing Mutations}
Mutations, random variations of existing solutions, represent one of the core concepts of evolutionary systems.
Over the life time of a system, mutations generate new candidate solutions based on the fittest solutions of a generation and replace solutions with worse performance.
As mutations are inherently random, we will also consider system architectures with so fine-tuned specialization in terms of the combined options that a normal monolithic system would not consider except for a rock solid case that becomes a major market need.

Despite the essential role of randomness, an evolutionary system can be flexible enough to take input 
from simple cost models in order  to provide a hybrid solution which also learns from the current state-of-the-art 
on data systems design or from knowledge that the research community develops in the future.
Cost models can give a rough indication on the space of mutations that can be first considered for a given query
or set of queries, minimizing the search space, 
and casting such hybrid random choice/deterministic model solutions as an attractive 
research path for quickly finding good solutions during the evolution process.

\paragraph{Updates and Streaming}
New data or updates arriving at rapid rates is one of the characteristics of many modern data-driven scenarios.
Evolutionary systems handle new data by placing it in the cold data section of the storage engine (Figure \ref{f:architecture}).
If relevant queries arrive, the respective storage modules will mutate the data to the appropriate format to match the 
required access pattern.    
Updates on existing and already mutated data need to affect the storage modules directly. 
Since such data may exist in multiple mutations, more than one copy\fxnote{copies} of the data may exist 
and may have to be updated, favoring solutions that incrementally and dynamically update data parts only if needed,
i.e., only if a relevant query arrives. 
Furthermore, the pace of evolution in specific data parts 
can adjust depending on the query and update frequency on these data parts.

\paragraph{The Cost of Evolution}
Enabling seamless transformations is the most 
critical challenge for evolutionary systems.  
Most times, transformation requires data replication and reorganization 
to enable the optimizations of the new architecture.
The challenge is to minimize, I/O costs, cache misses, CPU costs and memory footprint as well as
to exploit any opportunities for sharing across architectures.
For example, to create $N$ new data formats for a specific 
data partition, the system can minimize costs by reading this data only once 
through the memory hierarchy while producing all desirable formats in one go. 
In addition, data transitions do not need to happen in big monolithic steps.
A query can trigger a transition only for the data it touches,
leaving the task of transitioning the rest of the data to future queries.  
Thus, as shown in Figure \ref{f:architecture},
transformations are\fxnote{is} a central part of the execution engine itself, 
as opposed to becoming a component which is build on top of a traditional engine.

\section{Proof of Concept}
\label{sec:evaluation}
In this section, we discuss early results of our proof of concept of an evolutionary data system.
We show that the combination of mutations and natural selection 
is sufficient to continuously adapt a data system to meet the needs of a dynamic workload.

\paragraph{Storage and Access Modules}
Our evolutionary data system prototype contains storage and access modules that can 
mimic the behavior of a pure column-store, a key-value store as well as that of hybrids.
Each storage module is responsible for a horizontal part of the data and can, regarding storage patterns, mutate across the whole design space of the options above\fxnote{above options}. 
It may store data as key-value pairs or it may split the value part all the way to its individual properties
to mimic a pure column-store. In addition, it may split the value of each key-value pair 
into any combination of properties, effectively creating hybrid layouts. 
Access modules are tailored to access each storage module with the proper code depending on the storage layout
to minimize instruction misses and cache misses.

\paragraph{Mutations and Selection of the Fittest}
For every incoming query, the system chooses one active solution (combination of storage and access patterns) 
of the current population to answer the query. 
As more queries arrive, all active solutions in the current population will be tested.
The execution engine continuously observes the performance of the system on the incoming queries (response time) and, 
periodically, induces the evolutionary process as outlined in Algorithm \ref{alg:pseudocode};
In a process of natural selection, the system eliminates $X$\% of the solutions in the current population (Line 2), 
and generates a set of new random mutations of the surviving layouts (Line 3 + 4). 
Together, the mutations and the surviving layouts form the new population, i.e., the layouts and access patterns of the next generation.

\vspace{-.5em}
\begin{algorithm}
\begin{algorithmic}[1]
\small{
\Procedure{Evolve}{$current\_gen,perf\_stats,pop\_size$}%\Comment{The g.c.d. of a and b}
%\State $r\gets a\bmod b$
%\While{$r\not=0$}\Comment{We have the answer if r is 0}
\State $parents \gets \texttt{SelectParents}(current\_gen,perf\_stats)$ %\Comment{Select Population based on performance}
\State $next\_gen \gets \texttt{GeneratePopulation}(parents,pop\_size)$
\State $new\_gen \gets \texttt{Mutate}(parents,next\_gen)$
\State \textbf{return} $new\_gen$
%\State $b\gets r$
%\State $r\gets a\bmod b$
%\EndWhile\label{euclidendwhile}
%\State \textbf{return} $b$\Comment{The gcd is b}
\EndProcedure
}
\caption{High-Level \textit{Natural Selection} Algorithm}
\label{alg:pseudocode}
\end{algorithmic}
\end{algorithm}
\vspace{-1em}

\paragraph{Setup\fxnote{Set-up}}
We demonstrate results on a \testmachine.
The setup is as follows.
At first, we create 16GB of main-memory resident data that is initially generated and stored as key-value pairs, 
where the corresponding value part of each key contains 7 properties, stored as a consecutive payload.
All properties are initialized with random integer values.
Keys are stored as a separate column. %, while values are stored as a consecutive payload. 
 %are of integer type and contains random values. 
Queries access several key-value pairs and may be interested in all properties of a key-value pair 
or \fxnote{it may be interested }only in a subset of them.
Each query performs a series of scans and aggregations on one or more properties of each key-value pair.  
\fxnote{Regarding the evolution properties, i}In this example the system uses a fixed population size of 4 candidate solutions,
while in each phase, it eliminates 50\% of them and replaces them with random mutations of the surviving ones. 

\paragraph{Evolving Across Architectures}
Figure \ref{f:exp1} shows how our evolutionary prototype continuously  evolves to reach the optimal layout
when confronted with a series of evolving queries. 
Each phase of the query workload contains a random set of queries that touch  
a given set of combinations of properties of the key-value pairs.
In Phase 1, queries need all properties while in subsequent workload 
phases queries work on subsets of the properties in key-value pairs.
The queries have different selectivities, which trigger a variety of data layout decisions.
This means that there is an optimal 
storage layout and access pattern which contains physical representations of all combinations of 
properties as they are accessed in the query workload. 
The graph shows four curves (Candidate A-D), which represent the performance of the system (CPU cycles), as the set of solutions 
evolves through mutations, forming new generations ($x$-axis). 
%when using each candidate solution in the population. 
The solutions are ranked; Candidate A represents the fittest solution in each generation
while Candidate D shows the performance of the worst solution. 
Figure \ref{f:exp1} shows that starting from the original key-value storage and access pattern (\textit{Phase~1}), 
the system steadily evolves once the workload changes (as of \textit{Phase~2}), i.e., 
the fittest solution becomes continuously faster, eventually reaches
the optimal performance and then stays stable.
The optimal performance is taken by manually creating the perfect storage layout and access patterns. 
%When the workload changes the system adjusts its storage and access patterns. 
\fxnote{This} Adjustment of storage and access patterns is achieved by always trying out new mutations. \fxnote{when the workload changes the} Due to natural selection, new mutations that outperform the previous
fittest solution, will take its place and replace it in the next system generation\fxnote{and are the ones to survive}.

\paragraph{Summary}
Overall the experiment demonstrates a promising behavior to adapt the architecture and core components 
of a data system on-the-fly. Other than the brief analysis shown here, several more metrics and details  are of interest
which we omit in this version of the paper due to space restrictions, e.g, population size,
the number of solutions that are eliminated in each round, 
whether each single query runs on one or more candidate solutions in parallel, as well as selectivity 
and concurrent queries. 

\section{Challenges and Opportunities}
\label{sec:research-challenges}
\todo{Based on preliminary evaluation: Show that topic opens up a new research area. What are challenges that arise in the context of evolutionary DB systems? What are chances and interesting opportunities? Is there some kind of "grand vision"?}
Realizing the vision of evolutionary database systems opens up a vast 
research space and requires rethinking of how database systems are 
structured and components interact with one another.
In this section, we outline several areas for future investigations as well as opportunities to rethink existing concepts and solutions.

\paragraph{Tailored Architectures} 
Over the last decades, database systems research has yielded a heterogeneous set of 
data management models and techniques, ranging from key-value stores to highly complex transactional systems.
The main trend is that new models are treated as separate systems instead of different manifestations of shared components
even though the actual differences of these systems are not drastic.
One of the most important reasons for this trend is that a new system needs to support 
only the new functionality and design and implement it in a way that fits the new model exactly to get the best possible 
performance as opposed to a generic system that does many things well but none great.
Evolutionary systems bring the opportunity to create such tailored architectures in an autonomous way with minimal human input.
By being able to spin off and test numerous architectures that otherwise would be impossible 
(or extremely expensive and time consuming) to create manually, evolutionary systems represent 
a promising path to handle the deluge of data-driven applications with continuously evolving requirements. 
There is a plethora of research challenges in this direction including rich languages to describe 
modules, mutation rules and fitness metrics. In addition, developing standard APIs for modules to interact
and enabling seamless addition of new modules are\fxnote{is} needed to be able to easily enrich an existing system. 

\paragraph{Cloud-Computing and Distributed Databases}
Even though the basic concept of evolutionary database systems is applicable to nearly all types of runtime environments, cloud computing provides a particularly interesting setting.
As most cloud providers significantly overprovision their infrastructure in order to serve 
peak hours \cite{BreweHPTSKeynote}, idle resources during off-peak can be used to 
execute queries on multiple candidate solutions in parallel. 
This way alternative solutions can be directly compared with each other without affecting the overall query response time.\\
Another interesting research direction is to explore how evolutionary systems can be leveraged to dynamically scale databases across multiple nodes.
In this context, evolutionary concepts could form the bases for work distribution, fragmentation and allocation algorithms. Further, the inherited replication of the evolutionary approach brings data redundancy at no extra cost.

\paragraph{Adaptation to Hardware}
Evolutionary systems bring opportunities in dealing with diverse and evolving hardware.
An evolutionary system ranks a candidate architecture by its performance, i.e., by the end result, 
which allows testing of diverse solutions and making decisions based on performance as opposed to 
strictly based on complex tuning parameters. 
Given the inherent and increasing complexity of diverse hardware, evolutionary systems
can simplify matching of design solutions to hardware by being able 
to automatically test a large number of candidate solutions.
This property is especially important in cloud settings and other big infrastructures
with diverse hardware, where choosing a single architecture for the whole infrastructure
leads to suboptimal performance for part of the machines. An evolutionary system, on the other hand,
can choose a different solution
for each machine as it can spin off multiple different architectures. 

\todo{We need to extend the concepts of evolution to more database components such as: concurrency control, auxiliary data structure, query optimizers, distributed, benchmarking }

\paragraph{Evolutionary Database Kernels}
The design of evolutionary kernels provides numerous opportunities for innovation.
Essentially all components of database systems and all design decisions can be subject to evolution, i.e., 
testing multiple competing solutions and spinning off new combinations of design solutions continuously.
Fundamental and traditional features such as concurrency control, auxiliary data structures as well as  query plan generation
can all be reconsidered in the presence of an evolutionary engine.
For example, a query optimizer could automatically evolve its cost model to become more accurate or 
automatically develop rules to determine which set of replicated data should be used. 
In addition, important questions include how properties like \textit{population size} and \textit{selectivity} affect the speed and outcome of the evolutionary process and how this knowledge can be used to automatically configure evolutionary data systems for use on a specific machine.
Furthermore, given the drastically different architecture and feature set of evolutionary systems, new benchmarking techniques will
be required to stress test %the evolution process and 
the core components of such systems.

\newpage
\section{Related Work}
\label{sec:related-work}
Evolutionary systems draw inspiration from numerous areas and pioneering work in database architectures.
Here we position evolutionary systems with respect to this past work and highlight the new opportunities.
Due to space restrictions, we only refer to the most relevant areas in this version of the paper.  

\paragraph{Auto-tuning in Data Systems}
To ease the process of tuning and setting up data systems 
the research community came up with auto-tuning tools \cite{MS:ViewIndex}.
Most of the recent work focuses on offline tools or periodic online analysis that guide the physical design process. 
Auto-tuning in database systems can be considered as the first effort to adapt a 
database architecture in an automated way. Evolutionary systems extend these concepts 
in that they 
(a) evolve not only on local variations of a single architecture but across architectures and 
(b) provide a way to perform all those actions without external human control or design and without workload knowledge. 

\paragraph{Hybrid Systems}
A series of pioneering projects have proposed ideas where the storage layout 
blends properties from more than one micro-architectures  
resulting in hybrid systems \cite{FracturedMirrors,H2O, DataMorphing,Hyrise,pax}.
Past work mainly focuses to proposing hybrids between the two global relational layouts, 
column-stores and row-stores. Evolutionary systems push these ideas further by  bringing the opportunity to 
adapt and evolve across different architectures (not just column-stores and row-stores). \fxnote{and 
by}By relying on a fully automated evolutionary process they also bring the opportunity to
create systems that fit a new application exactly by testing numerous combinations of alternative architectures. 

\paragraph{Adaptive Data Systems Architectures}
Recent research has proposed solutions to overcome the problem of static, 
data-independent decisions, by providing several levels of workload adaptivity. 
Such work includes ideas that try to increase performance by dynamically optimizing query plans 
based on data properties \cite{eddies,contentrouting,multiroutestreams}, 
as well as by adaptively building auxiliary data structures such as indexes \cite{StratosCracking, Pirk2014, Idreos2007a, Petraki2015, Karras2016, Zoumpatianos2016, Halim2012, Graefe2010b, Idreos2009, Idreos2011, Graefe2014}
and techniques to adaptively load data from flat files \cite{NoDBcidr,Alagiannis2015}.
Another recent stream of work in adaptive processing is the idea of on-the-fly code generation 
of system components to fit the workload's data 
and requests exactly \cite{CompiledQueries} 
and to create kernels that adapt to provide interactive exploration \cite{dbTouchCIDR2013}.
In contrast to past work on adaptive database architectures, which has focused on solving local 
tuning problems within a single database architecture in order to cope with dynamic workloads, 
evolutionary systems push these ideas further by bringing the opportunity to adapt across 
several architectures as well as to synthesize new variations of architectures; 
they enable testing of more fine-grained combinations of the various system components, 
thus also raising new opportunities on how to design future data systems.

\paragraph{Multi-store Systems and Multi-purpose Engines}
Multiple approaches have been proposed to deal with the increasing system 
architectures diversity. % and to ease the process of tuning systems.
RodentStore \cite{RodentStore} envisions a hybrid system which lets users 
define data layouts by means of a declarative language.
The idea of LegoBase \cite{languagedb} takes this one step further by proposing that 
the whole engine should be designed in a declarative way.
Another stream of work proposes to create ``wrapper systems'' that operate multiple underlying systems in parallel
and navigate the queries and data to the proper system, e.g., \cite{miso,Odyssey,polybase,multiplex}, while another vision
suggests that data should be kept in a flexible format within a single system and materialize multiple views
depending on what the requirements of the application are \cite{DJ11}.
Finally, there has also been work on querying multiple systems that support different languages 
\cite{engineIndependenceLanguages}. % as well transferring scripts from one system to another.
 These lines of work are the most relevant to our vision\fxnote{ here} as they are all motivated by similar problems and goals.
They are complementary, though, to the approach they take and to the opportunities they bring. 
Evolutionary systems propose the unique vision of adopting the concept of evolution in database architectures
and bring the opportunity to  automatically synthesize tailored architectures with just enough functionality and footprint 
to match new applications.

\paragraph{Genetic Algorithms and Autonomous Systems}
In other areas of computer science, the concept of evolution has inspired pioneering work, 
such as evolutionary algorithms \cite{EvolutionaryAlgorithms} in artificial intelligence.
The concepts of evolution in database architectures bring several unique 
challenges that are intrinsic to the art of designing database architectures and data-driven application 
scenarios. In particular, when dealing with data and access methods in data-driven applications,
a system has to serve running queries as quickly as possible while any actions on the data can be very costly,
depending where the data lives in the memory hierarchy.

In fact, the vision of systems that can evolve has been 
as old as computer science itself.
John von Neumann wrote exciting texts about systems that can automatically create new systems
without human supervision \cite{SelfAutomata}.
We see our work as a stepping stone in this direction in the context of systems for data analytics.

\section{Summary}
\label{sec:conclusions}
Evolutionary systems represent a radically new way to think about data systems architectures where all design and optimization choices can be settled via an evolutionary process without human intervention.
This brings numerous opportunities towards systems that can be automatically fine-tuned to fit new and evolving application requirements and with just enough systems footprint.
A few first steps in this direction have happened for key-value stores \cite{Dayan2017}, ligthweight indexing \cite{Qin2016} and join processing \cite{Liu2016}.

\bibliographystyle{abbrv}
\bibliography{refs,../../../das/Bibliography-Mendeley/library-short,../../../das/Bibliography-Mendeley/to-be-added} 
\end{document}